\documentclass[fleqn,usenatbib]{mnras}

\usepackage[T1]{fontenc}
\usepackage{ae,aecompl}

\usepackage{graphicx}	
\usepackage{amsmath}	
\usepackage{amssymb}	
\usepackage{longtable}
\usepackage{array}
\usepackage{adjustbox}

\renewcommand\ion[2]{#1$\;${\small\rmfamily{#2}}\relax}
\newcommand\tinyion[2]{#1$\,${\tiny\rmfamily{#2}}\relax}

\usepackage{newtxtext,newtxmath}

\title[Chemical abundances in metal-poor HII regions]{Chemical abundances in 7 metal-poor HII regions and a determination of the primordial helium abundance}

\author[Valerdi, M. et al.]{
Mabel Valerdi,$^{1}$\thanks{E-mail: mvalerdi@astro.unam.mx}
Antonio Peimbert$^{1}$ \& Manuel Peimbert$^{1}$
\\
$^{1}$Instituto de Astronom\'ia, Universidad Nacional Aut\'onoma de M\'exico, Apdo. Postal 70-264 Ciudad Universitaria, M\'exico}

\date{Accepted XXX. Received YYY; in original form ZZZ}

\pubyear{2019}

\begin{document}
\label{firstpage}
\pagerange{\pageref{firstpage}--\pageref{lastpage}}
\maketitle

\begin{abstract}
We conducted a long-slit spectrophotometry analysis to obtain the chemical abundances of seven metal-poor \ion{H}{II} regions in three galaxies: UM~160, UM~420, and TOL~0513$-$393. The data have been taken with the Focal Reducer Low Dispersion Spectrograph 1 (FORS1) at the 8.2-m Very Large Telescope. We derived the physical conditions and the chemical abundances of N, O, Ne, S, Ar and Cl. We also performed a detailed analysis that involves abundance determinations using the $t^2$ formalism. Based on \ion{He}{I} recombination line intensity ratios, together with the {\tt Helio14} code, we derived the abundance of He. In addition, for a value $\Delta Y/\Delta Z_O \rm =3.3\pm 0.7$, we have estimated that the primordial helium abundance by mass is $Y_{\rm P}=0.2448\pm0.0033$. This value agrees with values derived from Standard Big Bang Nucleosynthesis and with other recent determinations of $Y_{\rm P}$.
\end{abstract}




\section{Introduction}

The knowledge of the chemical composition of \ion{H}{II} regions is critical to our current understanding of the chemical composition of the Universe; precise determinations are of great importance, since they provide observational restrictions to galactic chemical models. Examples of these are: the determination of radial abundance gradients in spiral galaxies; together with other physical properties, such as mass-luminosity relation, we can study the mass-metallicity ratio for spiral galaxies, or, with stellar evolution models, we can study the chemical evolution of galaxies \citep[e.g.,][]{Tremonti2004,PerezM2008,Liu2008,Peimbert2011,Izotov2015,Pilyugin2017,Carigi2019}. Spectra obtained from \ion{H}{II} regions contain emission lines of different ions, which belong to elements present in the ionized gas. H and He are studied by means of recombination lines (RLs), while most of the heavy elements are studied through collisionally excited lines (CELs). Typically, CELs have been used to derive heavy element abundances in \ion{H}{II} regions, because they are brighter and easier to detect than RLs; however, abundances obtained through RLs for heavy elements are less sensitive to temperature and density, than abundances obtained from CELs \citep{Tsamis2003,Esteban2009}.

A known and unresolved problem in the astrophysics of ionized nebulae relates to the discrepancy between abundances derived from RLs and those derived from CELs (while assuming constant temperature); the ratio between these abundances is called the abundance discrepancy factor \citep[ADF][]{Garcia2007,Garcia2013}, where abundances determined from RLs are always higher than those determined from CELs. For \ion{H}{II} regions the ADF is usually between 1.5 and 3 \citep[e.g.][]{Garcia2007,Toribio2017}. Several authors have tried to explain this discrepancy through different hypotheses \citep[e.g.,][]{Peimbert1969,Tsamis2005,Nicholls2012,Peimbert-Gloria2017}, but the debate is still open.

\citet{Peimbert1967} and \citet{Peimbert1969} first developed the formalism of temperature inhomogeneities ($t^2$), to determine chemical abundances. This formalism refers to the possibility of including thermal inhomogeneities in studies of photoionized regions. The thermal inhomogeneities usually have a standard deviation of approximately 20\% of the electron temperature ($\sigma_{T} \approx 20\% T_e$), but the $t^2$ value is different for each object, however this seemingly small variation can lead us to underestimate the chemical abundances derived from CELs by a factor of approximately 2 \citep[e.g.][]{Garcia2007,Esteban2018}. Unfortunately it is not possible to determine a $t^2$ for all objects (for example, \ion{H}{II} galaxies with high redshift or low intrinsic brightness), and in these cases it is possible to use an average $\left<t^2\right>$ value, obtained from those \ion{H}{II} regions where it is possible to determine $t^2$ \citep{Peimbert2012}.

Studies of abundances in extragalactic \ion{H}{II} regions find heavy element abundances that range from $0.02$ \({\rm Z}_\odot\) \citep[e.g.,][]{Izotov2007, Izotov2011, Peimbert2002, Peimbert2012} all the way up to several times the solar abundance \citep{Toribio2017,Esteban2020}. To find low metallicity \ion{H}{II} regions we need to look at the outer parts of spiral arms of late-type galaxies, or better yet at irregular-type galaxies; in these regions, the process of enrichment in heavy elements by stellar nucleosynthesis has been limited and this provides an opportunity for the study of the primordial chemical composition \citep{Izotov2007,Peimbert2002,Peimbert2012}. Furthermore, these regions have undergone little evolution in their helium abundance, this has the consequence that the He/H abundance should be close to the primordial value and therefore help provide restrictions for cosmological theories \citep{Porter2009,Izotov2014,Peimbert2016}. The study of metal-poor \ion{H}{II} regions allows us to understand some astrophysical processes that occur in the early Universe, which would otherwise be inaccessible for observation \citep[e.g.,][]{Kunth2000,Kniazev2003,Pustilnik2007,Brown2008}. We are interested in objects that have an oxygen abundance smaller or similar to $\sim0.1$ \({\rm Z}_\odot\) solar, equivalent to $12+\log({\rm O}/{\rm H})\lesssim7.65$.

\citet{Torres1974} were the first to use observations of  metal-poor \ion{H}{II} regions to determine the primordial helium abundance ($Y_{\rm P}$). Many other studies have been done since, improving on this first determination and showing the difficulties in having an accurate measurement \citep[e.g.,][]{Olive2001,Olive2004,Izotov2007,Peimbert2007}. A historical review on the determination of the primordial helium abundance was presented by \citet{Manuel2008}. We currently have the desire to obtain determinations with accuracy better than 1\%, since a highly accurate determination plays an important role to understand the Universe. In particular, it is desirable to be able to restrict: the Big Bang Nucleosynthesis models, the physics of elementary particles, and the study of galactic chemical evolution.

Other recent determinations of the primordial helium abundances using \ion{H}{II} regions have been done by \citet{Izotov2014, Aver2015, Peimbert2016, Valerdi2019, Vital2019,Aver2020, Hsy2020,Kurichin2021}.

In Section 2 we present the observations, as well as a description of the reduction procedure. In Section 3 we show the line intensity corrections due to extinction and underlying absorption. The determination of physical conditions and of chemical abundances, ionic and total, are shown in Sections 4 and 5. Finally in Sections 6 and 7 we present the primordial abundance determination and the conclusions.

\section{Observations}

Long-slit spectra were obtained with the Focal Reducer Low Dispersion Spectrograph 1, FORS1 at  the Very Large Telescope Facility (VLT) in Chile. The slit is $410''$ long and was set to be $0.51''$ wide, for a spectral resolution of $\sim$1500. We used three grism settings: GRIS-600B+12, GRIS-600R+14 with filter GG435, and GRIS-300V with filter GG375. Regardless of the air mass value, the linear atmospheric dispersion corrector was used to keep the same observed region within the slit. Table \ref{tab:grism} shows the resolutions and wavelength coverage for the emission lines observed with each grism.

\begin{table}
\centering
\caption{Observation Settings.}
\label{tab:grism}
\begin{tabular}{lcccc}
\hline
Grism & Filter & $\lambda$ & Resolution & Exp. time\\
  & & (\AA) & ($\lambda$/$\Delta\lambda$) & (s)\\ \hline
600B+12 & $-$ & 3450$-$5900 & 1300 & $720 \times 3$ \\
600R+14 & GG435 & 5350$-$7450 & 1700 & $600 \times 3$ \\
300V & GG375 & 3850$-$8800 & 700 & $120 \times 3$ \\
\hline
\end{tabular}
\end{table}

We observed 7 metal-poor \ion{H}{II} regions in 3 galaxies: UM~160, UM~420, and TOL~0513$-$393.

\begin{figure}
\includegraphics[width=\columnwidth]{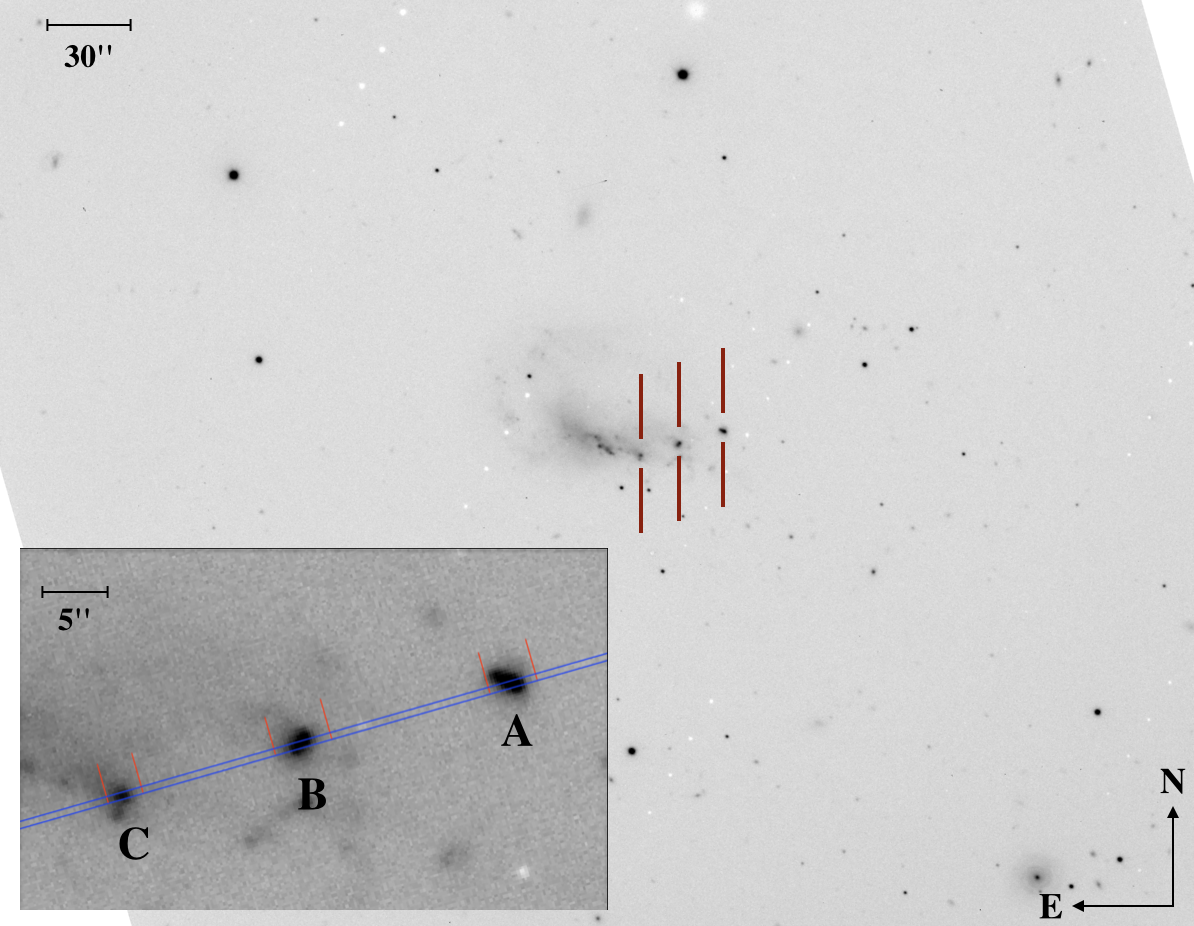}
\caption{Image of UM~160 obtained with VLT. The lower left image shows the object, where the red lines mark the 18, 21, and 13 pixels corresponding to the sections A, B, and C respectively; the slit is indicated by the two blue lines.}
\label{fig:160}
\end{figure}

For galaxy UM~160 ($\alpha=03^{h}24^{m}23^{s}.1$, $\delta=-00^{\circ}06'29''$) the slit was placed at $-106^{\circ}.5$ with respect to North (see Figure \ref{fig:160}). In it, 3 ionized regions are clearly visible (i.e. 3 \ion{H}{II} regions). Each of the \ion{H}{II} regions was analyzed independently: it is obvious that each region will have, its own degree of ionization, its own ionization zone, potentially different chemical compositions, and different temperatures. Region A is 18 pixels ($\sim3''.6$) long, region B is 62 pixels ESE of region A and has a length of 21 pixels ($\sim4''.2$), and region C is 52 pixels ESE of region B and has a length of 13 pixels ($\sim2''.6$).

\begin{figure}
\includegraphics[width=\columnwidth]{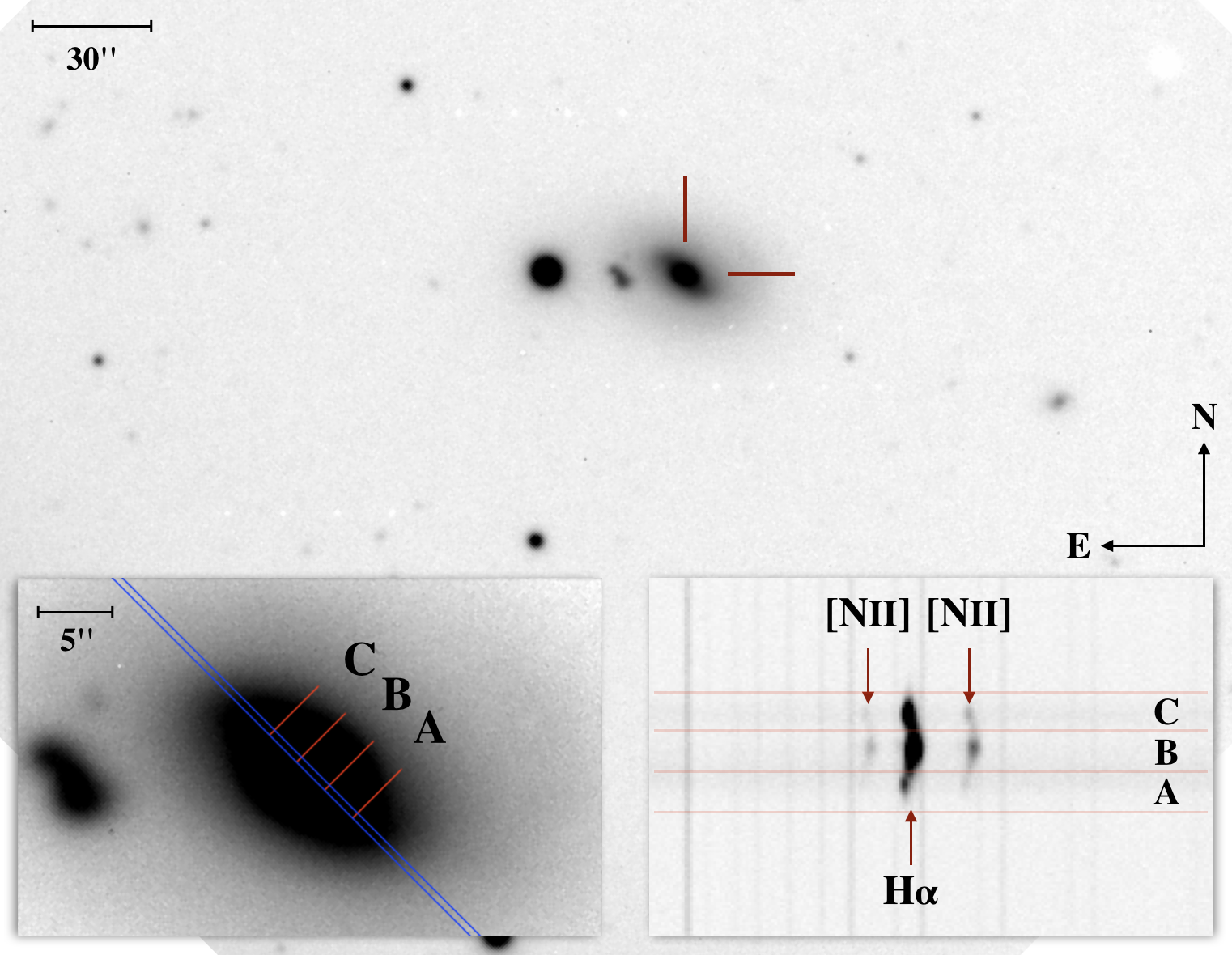}
\caption{Image of UM~420 obtained with VLT. The lower left image shows the object, where the red lines mark the 14, 15, and 15 pixels corresponding to the regions A, B, and C respectively; the slit is indicated by the two blue lines. The lower right image shows the H$\alpha$ $\lambda4861$, [\ion{N}{II}] $\lambda\lambda 6548,6583$ lines in the three 3 sections analyzed.}
\label{fig:420}
\end{figure}

For galaxy UM~420 ($\alpha=02^{h}20^{m}54^{s}.5$, $\delta=00^{\circ}33'24''$) the slit was placed at $-45^{\circ}.0$ with respect to North. Figure \ref{fig:420} shows the direct image, it appears to be a single  \ion{H}{II} region. The lower right image shows a position velocity diagram that includins H$\alpha$, $\lambda6548$[\ion{N}{II}], and $\lambda6583$[\ion{N}{II}], where it is clear that there are 3 components with different velocities.
Using the same reasoning, that we used to divide the photons of UM~160 into 3 regions, again we decided to divide the photons of UM~420 into 3 regions; fortunately the different velocity components are also separated in space (mostly). We defined region A to be 14 pixels ($\sim2''.8$) long and is NE of the center of the galaxy, we defined region B to be 15 pixels ($\sim3''.0$) and corresponds to the center of the galaxy, and region C to be 15 pixels ($\sim3''.0$) long and is SW of the center of the galaxy; all 3 regions directly adjacent to one another.

\begin{figure}
\includegraphics[width=\columnwidth]{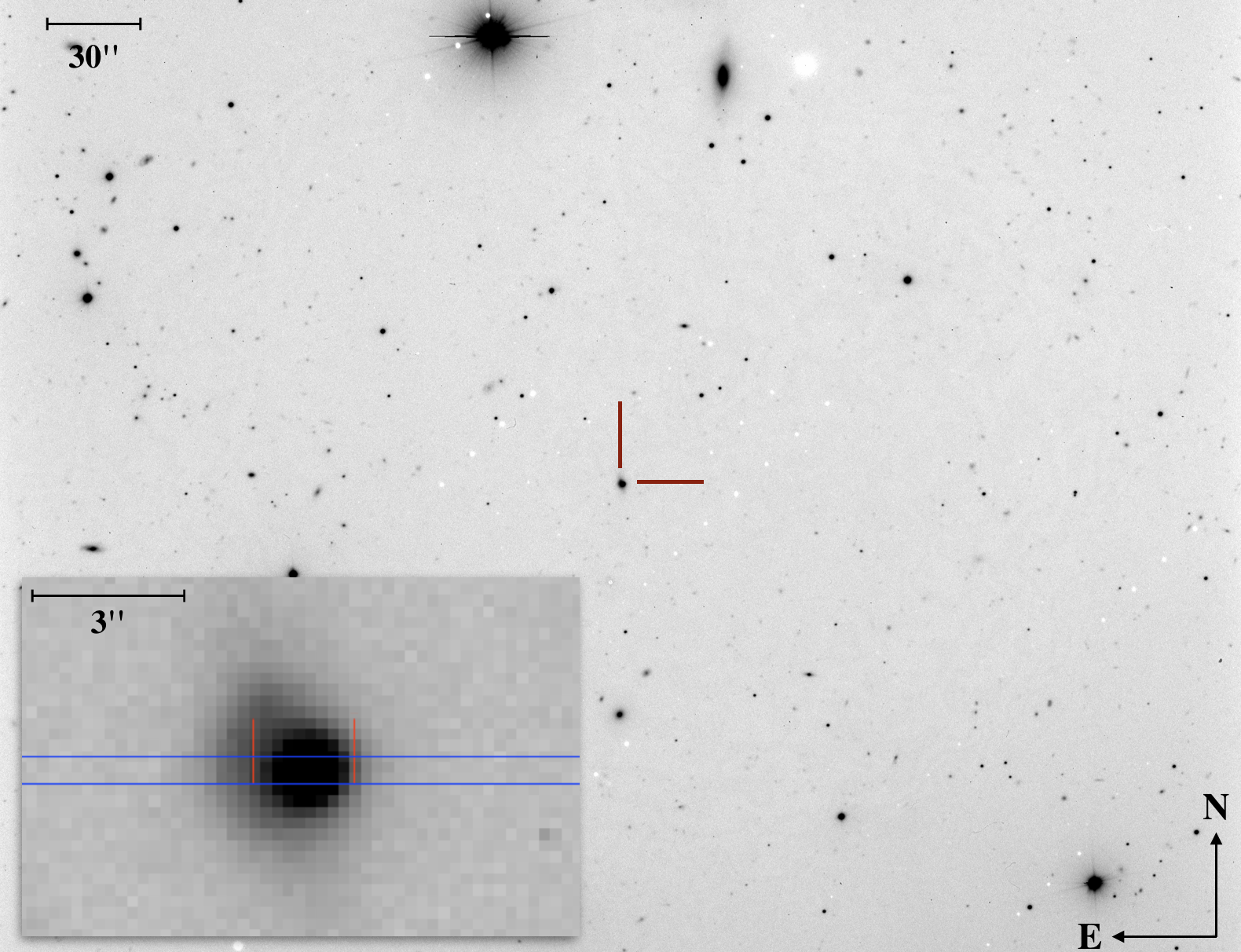}
\caption{Image of TOL~0513$-$393 obtained with VLT. The lower left image shows the object, where the red line mark the 9 pixels corresponding to section analyzed; the slit is indicated by the two blue lines.}
\label{fig:0513}
\end{figure}

Finally for galaxy TOL~0513$-$393 ($\alpha=05^{h}15^{m}19^{s}.8$, $\delta=-39^{\circ}17'41''$) the slit was placed at $-90^{\circ}.0$ with respect to North (see Figure \ref{fig:0513}). In this galaxy, there is no evidence to suggest multiple \ion{H}{II} regions (and no useful way to separate the observed photons) so we assumed it is a single 9 pixels ($\sim1''.8$) long \ion{H}{II} region.

The spectra were reduced using \texttt{IRAF}\footnote{\texttt{IRAF} was distributed by National Optical Astronomy Observatories, which was operated by the Association of Universities for Research in Astronomy, under cooperative agreement with the National Science Foundation.}, following the standard procedure: bias subtraction, aperture extraction, flat field, and wavelength calibration. We also corrected effects caused by the medium between the observer and the object: redshift, underlying absorption and reddening. For the flux calibration we have used the following standard stars: LTT~2415, LTT~7389, LTT~7987, and EG~21 \citep{Hamuy1992,Hamuy1994}. We obtained three spectra for each analyzed region; blue spectra, red spectra, and low-resolution spectra see Table \ref{tab:grism}.  Each spectrum was taken in 3 integrations in order to eliminate cosmic rays; the total integration time for blue spectra was 36 minutes, for red spectra, was 30 minutes, and for low dispersion, it had a total time of only 6 minutes: we used the low-resolution spectra to connect the calibration between the red and blue spectra.

\section{Line intensities and reddening correction}

To measure the emission line fluxes, we used the \texttt{SPLOT} routine of the \texttt{IRAF} package. We have measured the flux in the line by integrating all the flux between two given limits and fitted local continuum which was estimated by eye. When two or more lines are blended, for example [\ion{O}{II}] $\lambda\lambda3726+29$, we deblended them by fitting Gaussian profiles with the same widths and thus estimated the individual fluxes.

To correct the fluxes for reddening we used the extinction law of \citet{Seaton1979}, with the reddening coefficient $c({\rm H}\beta)$, which is obtained by fitting the flux ratios of the brightest Balmer lines as compared with the theoretical values from the data by \citet{Storey1995}. In addition, we performed an underlying absorption correction by assuming that $EW_{\it abs}(\lambda)=EW_{\it abs}({\rm H}\beta) \times g(\lambda)$, where we used the $g(\lambda)$ values presented in \citet{Pena2012}; this correction allows us to eliminate the contribution of dust-scattered starlight; this correction is only necessary for H and He lines.
The reddening and the underlying absorption were fitted simultaneously using the following equation:

{\scriptsize
\begin{table*}
\centering
\caption{List of emission line intensities for all \ion{H}{II} regions. \label{tab:lines1}}
\begin{tabular}{lc@{}r@{\hspace{12pt}}r@{\hspace{5pt}}r@{\hspace{2pt}}c@{\hspace{10pt}}r@{\hspace{5pt}}r@{\hspace{2pt}}c@{\hspace{10pt}}r@{\hspace{5pt}}r@{\hspace{2pt}}c@{\hspace{10pt}}r@{\hspace{5pt}}r@{\hspace{2pt}}c}
\hline
\multicolumn{1}{l}{$\lambda$} &
\multicolumn{1}{c}{ID} &
\multicolumn{1}{c}{$f(\lambda)$} &
\multicolumn{1}{c}{$F(\lambda$)} &
\multicolumn{1}{c}{$I(\lambda$)} &
\multicolumn{1}{c}{\% Error}     &
\multicolumn{1}{c}{$F(\lambda)$} &
\multicolumn{1}{c}{$I(\lambda)$} &
\multicolumn{1}{c}{\% Error} &
\multicolumn{1}{c}{$F(\lambda$)} &
\multicolumn{1}{c}{$I(\lambda$)} &
\multicolumn{1}{c}{\% Error}     &
\multicolumn{1}{c}{$F(\lambda)$} &
\multicolumn{1}{c}{$I(\lambda)$} &
\multicolumn{1}{c}{\% Error} \\
\hline
 & & &\multicolumn{3}{c}{UM~160~A} &\multicolumn{3}{c}{UM~160~B}& \multicolumn{3}{c}{UM~160~C} &\multicolumn{3}{c}{UM~420~A} \\
\hline
3669	&	\tinyion{H}{I}	&	0.269	&	0.62	&	0.92	&	25	&	1.14	&	1.29	&	20	&	0.72	&	1.07	&	20	&	$\cdots$	&	$\cdots$	&	$\cdots$ 	\\
3671	&	\tinyion{H}{I}	&	0.269	&	$\cdots$	&	$\cdots$	&	$\cdots$	&	1.01	&	1.14	&	20	&	0.40	&	0.64	&	25	&	$\cdots$	&	$\cdots$	&	$\cdots$ 	\\
3674	&	\tinyion{H}{I}	&	0.269	&	$\cdots$	&	$\cdots$	&	$\cdots$	&	1.19	&	1.35	&	20	&	1.32	&	1.92	&	15	&	$\cdots$	&	$\cdots$	&	$\cdots$ 	\\
3676	&	\tinyion{H}{I}	&	0.269	&	$\cdots$	&	$\cdots$	&	$\cdots$	&	1.63	&	1.84	&	15	&	$\cdots$	&	$\cdots$	&	$\cdots$	&	$\cdots$	&	$\cdots$	&	$\cdots$	\\
3679	&	\tinyion{H}{I}	&	0.268	&	$\cdots$	&	$\cdots$	&	$\cdots$	&	0.81	&	0.96	&	20	&	1.07	&	1.63	&	20	&	$\cdots$	&	$\cdots$	&	$\cdots$ 	\\
3683	&	\tinyion{H}{I}	&	0.267	&	$\cdots$	&	$\cdots$	&	$\cdots$	&	1.86	&	2.13	&	15	&	1.26	&	1.94	&	20	&	$\cdots$	&	$\cdots$	&	$\cdots$ 	\\
3687	&	\tinyion{H}{I}	&	0.266	&	0.73	&	1.12	&	25	&	1.40	&	1.66	&	20	&	0.85	&	1.43	&	20	&	$\cdots$	&	$\cdots$	&	$\cdots$ 	\\
3691	&	\tinyion{H}{I}	&	0.265	&	0.72	&	1.13	&	25	&	2.27	&	2.64	&	15	&	0.75	&	1.36	&	20	&	$\cdots$	&	$\cdots$	&	$\cdots$ 	\\
3697	&	\tinyion{H}{I}	&	0.264	&	1.09	&	1.68	&	20	&	3.14	&	3.63	&	12	&	2.03	&	3.18	&	12	&	$\cdots$	&	$\cdots$	&	$\cdots$ 	\\
3704	&	\tinyion{H}{I}	&	0.262	&	1.14	&	1.78	&	20	&	2.78	&	3.28	&	12	&	2.59	&	4.05	&	12	&	$\cdots$	&	$\cdots$	&	$\cdots$ 	\\
3712	&	\tinyion{H}{I}	&	0.260	&	1.02	&	1.63	&	20	&	2.46	&	2.99	&	15	&	2.09	&	3.48	&	12	&	$\cdots$	&	$\cdots$	&	$\cdots$ 	\\
3722	&	\tinyion{H}{I}	&	0.257	&	1.63	&	2.53	&	15	&	2.92	&	3.55	&	12	&	2.64	&	4.34	&	12	&	$\cdots$	&	$\cdots$	&	$\cdots$ 	\\
3726	&	[\tinyion{O}{II}]	&	0.256	&	36.57	&	52.16	&	4	&	55.88	&	60.75	&	3	&	53.21	&	72.24	&	3	&	120.83	&	131.34	&	8	\\
3729	&	[\tinyion{O}{II}]	&	0.255	&	56.88	&	81.00	&	3	&	81.12	&	88.14	&	3	&	80.73	&	109.47	&	3	&	192.19	&	202.50	&	6	\\
3734	&	\tinyion{H}{I}	&	0.254	&	1.91	&	2.95	&	15	&	3.50	&	4.24	&	12	&	2.47	&	4.21	&	12	&	$\cdots$	&	$\cdots$	&	$\cdots$ 	\\
3750	&	\tinyion{H}{I}	&	0.250	&	1.64	&	2.69	&	15	&	1.32	&	2.35	&	20	&	0.58	&	2.30	&	25	&	$\cdots$	&	$\cdots$	&	$\cdots$  	\\
3771	&	\tinyion{H}{I}	&	0.245	&	2.19	&	3.55	&	15	&	3.07	&	4.52	&	15	&	1.53	&	4.05	&	15	&	$\cdots$	&	$\cdots$	&	$\cdots$  	\\
3798	&	\tinyion{H}{I}	&	0.238	&	3.29	&	5.14	&	12	&	3.23	&	4.99	&	12	&	2.10	&	5.32	&	12	&	$\cdots$	&	$\cdots$	&	$\cdots$  	\\
3819	&	\tinyion{He}{I}	&	0.233	&	1.09	&	1.60	&	20	&	0.48	&	0.76	&	25	&	1.38	&	2.23	&	15	&	$\cdots$	&	$\cdots$	&	$\cdots$ 	\\
3835	&	\tinyion{H}{I}	&	0.229	&	4.69	&	7.09	&	10	&	6.43	&	8.58	&	10	&	3.48	&	7.85	&	10	&	$\cdots$	&	$\cdots$	&	$\cdots$  	\\
3869	&	[\tinyion{Ne}{III}]	&	0.222	&	27.34	&	37.11	&	4	&	35.83	&	38.34	&	4	&	24.12	&	31.27	&	4	&	38.70	&	50.75	&	12	\\
3889	&	\tinyion{H}{I}$+$\tinyion{He}{I}	&	0.218	&	13.20	&	18.72	&	6	&	16.25	&	19.38	&	6	&	11.98	&	19.49	&	6	&	12.14	&	23.89	&	20	\\
3969	&	\tinyion{H}{I}$+$[\tinyion{Ne}{III}]	&	0.201	&	20.62	&	28.08	&	6	&	24.87	&	28.65	&	4	&	16.77	&	25.57	&	6	&	9.05	&	22.25	&	25	\\
4026	&	\tinyion{He}{I}	&	0.190	&	0.95	&	1.32	&	20	&	1.34	&	1.65	&	20	&	1.72	&	2.57	&	15	&	$\cdots$	&	$\cdots$	&	$\cdots$  	\\
4069	&	[\tinyion{S}{II}]	&	0.182	&	1.16	&	1.49	&	20	&	1.20	&	1.26	&	20	&	1.68	&	2.07	&	15	&	$\cdots$	&	$\cdots$	&	$\cdots$  	\\
4076	&	[\tinyion{S}{II}]	&	0.181	&	0.69	&	0.89	&	25	&	0.46	&	0.48	&	25	&	$\cdots$	&	$\cdots$	&	$\cdots$	&	$\cdots$	&	$\cdots$	&	$\cdots$	\\
4102	&	\tinyion{H}{I}	&	0.176	&	19.94	&	26.27	&	6	&	23.60	&	27.24	&	6	&	18.61	&	27.35	&	4	&	13.04	&	27.95	&	20	\\
4341	&	\tinyion{H}{I}	&	0.128	&	39.69	&	47.91	&	4	&	44.56	&	47.74	&	3	&	38.95	&	48.08	&	3	&	34.88	&	46.82	&	12	\\
4363	&	[\tinyion{O}{III}]	&	0.122	&	7.28	&	8.54	&	8	&	6.47	&	6.62	&	10	&	3.75	&	4.25	&	10	&	$\cdots$	&	$\cdots$	&	$\cdots$  	\\
4472	&	\tinyion{He}{I}	&	0.094	&	3.61	&	4.21	&	10	&	3.57	&	3.97	&	12	&	3.30	&	4.21	&	10	&	$\cdots$	&	$\cdots$	&	$\cdots$  	\\
4658	&	[\tinyion{Fe}{III}]	&	0.047	&	0.70	&	0.73	&	25	&	0.98	&	0.97	&	20	&	1.09	&	1.11	&	20	&	$\cdots$	&	$\cdots$	&	$\cdots$  	\\
4686	&	\tinyion{He}{II}	&	0.041	&	0.46	&	0.48	&	25	&	2.41	&	2.38	&	15	&	$\cdots$	&	$\cdots$	&	$\cdots$	&	$\cdots$	&	$\cdots$	&	$\cdots$	\\
4712	&	[\tinyion{Ar}{IV}]$+$\tinyion{He}{I}	&	0.034	&	0.90	&	0.93	&	25	&	0.77	&	0.76	&	25	&	0.66	&	0.66	&	25	&	$\cdots$	&	$\cdots$	&	$\cdots$  	\\
4861	&	\tinyion{H}{I}	&	0	&	100.00	&	99.00	&	2	&	100.00	&	98.30	&	2	&	100.00	&	98.50	&	2	&	100.00	&	99.70	&	6	\\
4922	&	\tinyion{He}{I}	&	$-$0.013	&	0.97	&	1.00	&	20	&	1.17	&	1.30	&	20	&	1.11	&	1.31	&	20	&	$\cdots$	&	$\cdots$	&	$\cdots$  	\\
4959	&	[\tinyion{O}{III}]	&	$-$0.021	&	184.69	&	176.24	&	2	&	167.74	&	160.51	&	2	&	142.42	&	132.74	&	2	&	119.60	&	116.02	&	6	\\
4986	&	[\tinyion{Fe}{III}]	&	$-$0.027	&	1.01	&	0.95	&	20	&	0.82	&	0.78	&	25	&	1.02	&	0.94	&	10	&	$\cdots$	&	$\cdots$	&	$\cdots$ 	\\
5007	&	[\tinyion{O}{III}]	&	$-$0.032	&	558.89	&	524.88	&	1	&	498.57	&	474.66	&	1	&	429.11	&	394.00	&	1	&	345.52	&	312.92	&	4	\\
5016	&	\tinyion{He}{I}	&	$-$0.034	&	2.67	&	2.56	&	15	&	2.51	&	2.56	&	15	&	2.12	&	2.20	&	12	&	$\cdots$	&	$\cdots$	&	$\cdots$ 	\\
5199	&	[\tinyion{N}{I}]	&	$-$0.074	&	0.70	&	0.62	&	25	&	$\cdots$	&	$\cdots$	&	$\cdots$ 	&	0.81	&	0.70	&	20	&	$\cdots$	&	$\cdots$	&	$\cdots$ 	\\
5270	&	[\tinyion{Fe}{III}]	&	$-$0.089	&	$\cdots$	&	$\cdots$	&	$\cdots$	&	0.43	&	0.40	&	25	&	$\cdots$	&	$\cdots$	&	$\cdots$	&	$\cdots$	&	$\cdots$	&	$\cdots$	\\
5518	&	[\tinyion{Cl}{III}]	&	$-$0.140	&	$\cdots$	&	$\cdots$	&	$\cdots$	&	0.55	&	0.50	&	25	&	0.60	&	0.47	&	25	&	$\cdots$	&	$\cdots$	&	$\cdots$  	\\
5538	&	[\tinyion{Cl}{III}]	&	$-$0.144	&	$\cdots$	&	$\cdots$	&	$\cdots$	&	$\cdots$	&	$\cdots$	&	$\cdots$	&	0.32	&	0.25	&	25	&	$\cdots$	&	$\cdots$	&	$\cdots$  	\\
5876	&	\tinyion{He}{I}	&	$-$0.216	&	13.90	&	10.05	&	6	&	10.83	&	9.64	&	8	&	13.91	&	10.16	&	6	&	18.20	&	15.80	&	15	\\
6301	&	[\tinyion{O}{I}]	&	$-$0.285	&	3.94	&	2.57	&	10	&	3.28	&	2.78	&	12	&	4.99	&	3.25	&	8	&	9.37	&	15.83	&	25	\\
6312	&	[\tinyion{S}{III}]	&	$-$0.286	&	2.41	&	1.57	&	15	&	1.52	&	1.29	&	20	&	2.15	&	1.40	&	12	&	$\cdots$	&	$\cdots$	&	$\cdots$  	\\
6364	&	[\tinyion{O}{I}]	&	$-$0.294	&	1.71	&	1.10	&	15	&	0.92	&	0.77	&	25	&	1.49	&	0.96	&	15	&	$\cdots$	&	$\cdots$	&	$\cdots$  	\\
6548	&	[\tinyion{N}{II}]	&	$-$0.320	&	2.74	&	1.69	&	12	&	3.18	&	2.65	&	12	&	6.01	&	3.73	&	8	&	18.70	&	21.79	&	15	\\
6563	&	\tinyion{H}{I}	&	$-$0.322	&	454.82	&	280.75	&	1	&	333.49	&	278.80	&	2	&	445.78	&	277.09	&	1	&	365.57	&	286.40	&	4	\\
6583	&	[\tinyion{N}{II}]	&	$-$0.324	&	7.85	&	4.83	&	8	&	8.60	&	7.16	&	8	&	15.29	&	9.44	&	6	&	44.74	&	41.79	&	12	\\
6678	&	\tinyion{He}{I}	&	$-$0.337	&	4.90	&	2.98	&	10	&	3.83	&	3.26	&	10	&	4.48	&	2.81	&	10	&	$\cdots$	&	$\cdots$	&	$\cdots$  	\\
6716	&	[\tinyion{S}{II}]	&	$-$0.342	&	18.65	&	11.17	&	6	&	18.48	&	15.26	&	6	&	31.42	&	18.93	&	3	&	58.42	&	52.00	&	10	\\
6731	&	[\tinyion{S}{II}]	&	$-$0.343	&	14.28	&	8.54	&	6	&	12.26	&	10.12	&	6	&	22.63	&	13.62	&	4	&	41.23	&	38.93	&	10	\\
7136	&	[\tinyion{Ar}{III}]	&	$-$0.391	&	9.96	&	5.56	&	8	&	7.61	&	6.14	&	8	&	11.18	&	6.30	&	6	&	$\cdots$	&	$\cdots$	&	$\cdots$  	\\
7282	&	\tinyion{He}{I}	&	$-$0.406	&	1.37	&	0.75	&	20	&	$\cdots$	&	$\cdots$	&	$\cdots$	&	$\cdots$	&	$\cdots$	&	$\cdots$	&	$\cdots$	&	$\cdots$	&	$\cdots$	\\
7320	&	[\tinyion{O}{II}]	&	$-$0.410	&	3.47	&	1.88	&	12	&	2.68	&	2.15	&	15	&	4.16	&	2.29	&	10	&	$\cdots$	&	$\cdots$	&	$\cdots$ 	\\
7330	&	[\tinyion{O}{II}]	&	$-$0.411	&	2.91	&	1.58	&	12	&	2.08	&	1.67	&	15	&	3.17	&	1.74	&	10	&	$\cdots$	&	$\cdots$	&	$\cdots$ 	\\
7753	&	[\tinyion{Ar}{III}]	&	$-$0.452	&	3.83	&	1.95	&	10	&	3.41	&	2.68	&	12	&	3.70	&	1.92	&	10	&	$\cdots$	&	$\cdots$	&	$\cdots$ 	\\
\hline
\end{tabular}
\end{table*}
}

{\scriptsize
\begin{table*}
\centering
\caption{List of emission line intensities for all \ion{H}{II} regions. \label{tab:lines2}}
\begin{tabular}{lc@{}r@{\hspace{12pt}}r@{\hspace{5pt}}r@{\hspace{2pt}}c@{\hspace{10pt}}r@{\hspace{5pt}}r@{\hspace{2pt}}c@{\hspace{10pt}}r@{\hspace{5pt}}r@{\hspace{2pt}}c}
\hline
\multicolumn{1}{l}{$\lambda$}    &
\multicolumn{1}{c}{ID}           &
\multicolumn{1}{c}{$f(\lambda)$} &
\multicolumn{1}{c}{$F(\lambda$)} &
\multicolumn{1}{c}{$I(\lambda$)} &
\multicolumn{1}{c}{\% Error}     &
\multicolumn{1}{c}{$F(\lambda)$} &
\multicolumn{1}{c}{$I(\lambda)$} &
\multicolumn{1}{c}{\% Error}     &
\multicolumn{1}{c}{$F(\lambda$)} &
\multicolumn{1}{c}{$I(\lambda$)} &
\multicolumn{1}{c}{\% Error}     \\
\hline
 & & &\multicolumn{3}{c}{UM~420~B} &\multicolumn{3}{c}{UM~420~C}&\multicolumn{3}{c}{TOL~0513$-$393} \\
\hline
3614	&	\tinyion{He}{I}	&	0.271	&	$\cdots$	&	$\cdots$	&	$\cdots$	&	$\cdots$	&	$\cdots$	&	$\cdots$	&	0.45	&	0.51	&	20	\\
3669	&	\tinyion{H}{I}	&	0.269	&	$\cdots$	&	$\cdots$	&	$\cdots$	&	$\cdots$	&	$\cdots$	&	$\cdots$	&	0.31	&	0.37	&	25	\\
3671	&	\tinyion{H}{I}	&	0.269	&	$\cdots$	&	$\cdots$	&	$\cdots$	&	$\cdots$	&	$\cdots$	&	$\cdots$	&	0.29	&	0.35	&	25	\\
3674	&	\tinyion{H}{I}	&	0.269	&	$\cdots$	&	$\cdots$	&	$\cdots$	&	$\cdots$	&	$\cdots$	&	$\cdots$	&	0.31	&	0.38	&	25	\\
3676	&	\tinyion{H}{I}	&	0.269	&	$\cdots$	&	$\cdots$	&	$\cdots$	&	$\cdots$	&	$\cdots$	&	$\cdots$	&	0.42	&	0.51	&	20	\\
3679	&	\tinyion{H}{I}	&	0.268	&	$\cdots$	&	$\cdots$	&	$\cdots$	&	$\cdots$	&	$\cdots$	&	$\cdots$	&	0.52	&	0.63	&	20	\\
3683	&	\tinyion{H}{I}	&	0.267	&	$\cdots$	&	$\cdots$	&	$\cdots$	&	$\cdots$	&	$\cdots$	&	$\cdots$	&	0.51	&	0.63	&	20	\\
3687	&	\tinyion{H}{I}	&	0.266	&	$\cdots$	&	$\cdots$	&	$\cdots$	&	$\cdots$	&	$\cdots$	&	$\cdots$	&	0.64	&	0.79	&	20	\\
3691	&	\tinyion{H}{I}	&	0.265	&	$\cdots$	&	$\cdots$	&	$\cdots$	&	$\cdots$	&	$\cdots$	&	$\cdots$	&	0.72	&	0.90&	15	\\
3697	&	\tinyion{H}{I}	&	0.264	&	$\cdots$	&	$\cdots$	&	$\cdots$	&	$\cdots$	&	$\cdots$	&	$\cdots$	&	0.90	&	1.13	&	15	\\
3704	&	\tinyion{H}{I}	&	0.262	&	0.90	&	1.53	&	25	&	$\cdots$	&	$\cdots$	&	$\cdots$ 	&	1.31	&	1.61	&	12	\\
3712	&	\tinyion{H}{I}	&	0.260	&	1.01	&	1.77	&	25	&	$\cdots$	&	$\cdots$	&	$\cdots$ 	&	1.25	&	1.58	&	12	\\
3722	&	\tinyion{H}{I}	&	0.257	&	$\cdots$	&	$\cdots$	&	$\cdots$	&	$\cdots$	&	$\cdots$	&	$\cdots$	&	2.28	&	2.77	&	10	\\
3726	&	[\tinyion{O}{II}]	&	0.256	&	90.93	&	112.08	&	3	&	75.85	&	86.00	&	4	&	21.34	&	23.85	&	3	\\
3729	&	[\tinyion{O}{II}]	&	0.255	&	118.57	&	145.34	&	3	&	110.92	&	124.51	&	4	&	26.83	&	29.98	&	3	\\
3734	&	\tinyion{H}{I}	&	0.254	&	$\cdots$	&	$\cdots$	&	$\cdots$	&	$\cdots$	&	$\cdots$	&	$\cdots$	&	2.03	&	2.53	&	10	\\
3750	&	\tinyion{H}{I}	&	0.250	&	1.54	&	2.77	&	20	&	$\cdots$	&	$\cdots$	&	$\cdots$ 	&	2.05	&	2.63	&	10	\\
3771	&	\tinyion{H}{I}	&	0.245	&	2.72	&	4.45	&	15	&	3.13	&	4.87	&	20	&	2.88	&	3.65	&	8	\\
3798	&	\tinyion{H}{I}	&	0.238	&	2.84	&	4.86	&	15	&	2.37	&	4.37	&	25	&	4.15	&	5.16	&	8	\\
3819	&	\tinyion{He}{I}	&	0.233	&	1.11	&	1.55	&	20	&	1.49	&	1.91	&	25	&	0.86	&	1.04	&	15	\\
3835	&	\tinyion{H}{I}	&	0.229	&	4.51	&	7.03	&	10	&	4.15	&	6.60	&	20	&	5.52	&	6.75	&	6	\\
3869	&	[\tinyion{Ne}{III}]	&	0.222	&	30.12	&	37.65	&	4	&	39.54	&	45.51	&	6	&	49.73	&	54.81	&	3	\\
3889	&	\tinyion{H}{I}+\tinyion{He}{I}	&	0.218	&	14.99	&	19.66	&	6	&	15.20	&	18.93	&	10	&	15.19	&	17.49	&	4	\\
3969	&	\tinyion{H}{I}+[\tinyion{Ne}{III}]	&	0.201	&	19.45	&	24.65	&	6	&	20.56	&	24.91	&	8	&	29.73	&	33.21	&	3	\\
4026	&	\tinyion{He}{I}	&	0.190	&	1.96	&	2.46	&	15	&	1.79	&	2.19	&	25	&	1.77	&	2.00	&	10	\\
4069	&	[\tinyion{S}{II}]	&	0.182	&	2.77	&	5.28	&	15	&	3.08	&	6.07	&	20	&	0.77	&	0.84	&	15	\\
4076	&	[\tinyion{S}{II}]	&	0.181	&	0.940	&	3.20	&	25	&	$\cdots$	&	$\cdots$	&	$\cdots$ 	&	0.27	&	0.29	&	25	\\
4102	&	\tinyion{H}{I}	&	0.176	&	21.00	&	26.04	&	6	&	22.03	&	26.34	&	8	&	22.349	&	24.95	&	3	\\
4341	&	\tinyion{H}{I}	&	0.128	&	41.31	&	47.00	&	4	&	43.14	&	47.22	&	6	&	42.84	&	46.11	&	2	\\
4363	&	[\tinyion{O}{III}]	&	0.122	&	3.87	&	6.11	&	12	&	7.97	&	10.63	&	12	&	14.59	&	15.42	&	4	\\
4472	&	\tinyion{He}{I}	&	0.094	&	3.48	&	4.01	&	12	&	4.12	&	4.63	&	20	&	3.76	&	4.04	&	7	\\
4658	&	[\tinyion{Fe}{III}]	&	0.047	&	1.16	&	2.90	&	20	&	1.69	&	3.99	&	25	&	0.41	&	0.42	&	20	\\
4686	&	\tinyion{He}{II}	&	0.041	&	0.64	&	2.35	&	25	&	$\cdots$	&	$\cdots$	&	$\cdots$ 	&	1.01	&	1.03	&	15	\\
4712	&	[\tinyion{Ar}{IV}]$+$\tinyion{He}{I}	&	0.034	&	$\cdots$	&	$\cdots$	&	$\cdots$	&	$\cdots$	&	$\cdots$	&	$\cdots$	&	2.50	&	2.55	&	10	\\
4741	&	[\tinyion{Ar}{IV}]	&	0.028	&	$\cdots$	&	$\cdots$	&	$\cdots$	&	$\cdots$	&	$\cdots$	&	$\cdots$	&	1.67	&	1.70	&	12	\\
4861	&	\tinyion{H}{I}	&	0	&	100.00	&	99.70	&	3	&	100.00	&	99.80	&	4	&	100.00	&	101.10	&	2	\\
4922	&	\tinyion{He}{I}	&	$-$0.013	&	1.02	&	1.15	&	25	&	$\cdots$	&	$\cdots$	&	$\cdots$ 	&	0.98	&	1.04	&	15	\\
4959	&	[\tinyion{O}{III}]	&	$-$0.021	&	148.37	&	144.73	&	2	&	169.61	&	166.21	&	3	&	258.13	&	257.21	&	1	\\
4986	&	[\tinyion{Fe}{III}]	&	$-$0.027	&	1.70	&	3.03	&	20	&	$\cdots$	&	$\cdots$	&	$\cdots$ 	&	0.43	&	0.43	&	20	\\
5007	&	[\tinyion{O}{III}]	&	$-$0.032	&	444.06	&	426.52	&	1	&	503.27	&	486.75	&	2	&	782.64	&	776.31	&	1	\\
5016	&	\tinyion{He}{I}	&	$-$0.034	&	$\cdots$	&	$\cdots$	&	$\cdots$	&	$\cdots$	&	$\cdots$	&	$\cdots$	&	1.56	&	1.63	&	12	\\
5199	&	[\tinyion{N}{I}]	&	$-$0.074	&	1.63	&	2.75	&	20	&	1.60	&	3.31	&	25	&	0.28	&	0.27	&	25	\\
5518	&	[\tinyion{Cl}{III}]	&	$-$0.140	&	$\cdots$	&	$\cdots$	&	$\cdots$	&	$\cdots$	&	$\cdots$	&	$\cdots$	&	0.36	&	0.34	&	25	\\
5538	&	[\tinyion{Cl}{III}]	&	$-$0.144	&	$\cdots$	&	$\cdots$	&	$\cdots$	&	$\cdots$	&	$\cdots$	&	$\cdots$	&	0.18	&	0.17	&	25	\\
5876	&	\tinyion{He}{I}	&	$-$0.216	&	13.52	&	11.31	&	6	&	12.05	&	10.86	&	10	&	13.89	&	12.83	&	4	\\
6301	&	[\tinyion{O}{I}]	&	$-$0.285	&	9.03	&	7.96	&	8	&	7.74	&	7.90	&	12	&	2.59	&	2.31	&	10	\\
6312	&	[\tinyion{S}{III}]	&	$-$0.286	&	2.50	&	2.85	&	15	&	2.04	&	3.00	&	25	&	1.72	&	1.53	&	12	\\
6364	&	[\tinyion{O}{I}]	&	$-$0.294	&	3.25	&	3.40	&	12	&	2.34	&	3.23	&	25	&	0.86	&	0.77	&	15	\\
6548	&	[\tinyion{N}{II}]	&	$-$0.320	&	9.91	&	8.44	&	8	&	5.55	&	5.83	&	15	&	1.34	&	1.18	&	12	\\
6563	&	\tinyion{H}{I}	&	$-$0.322	&	370.41	&	281.54	&	2	&	332.19	&	281.01	&	2	&	321.07	&	282.74	&	1	\\
6583	&	[\tinyion{N}{II}]	&	$-$0.324	&	31.28	&	24.59	&	4	&	16.21	&	14.80	&	10	&	5.13	&	4.51	&	6	\\
6678	&	\tinyion{He}{I}	&	$-$0.337	&	4.39	&	3.35	&	12	&	3.90	&	3.36	&	20	&	3.96	&	3.49	&	8	\\
6716	&	[\tinyion{S}{II}]	&	$-$0.342	&	32.50	&	25.00	&	4	&	25.80	&	22.65	&	7	&	7.19	&	6.27	&	6	\\
6731	&	[\tinyion{S}{II}]	&	$-$0.343	&	25.33	&	19.64	&	6	&	19.67	&	17.52	&	8	&	6.10	&	5.32	&	6	\\
7065	&	\tinyion{He}{I}	&	$-$0.383	&	3.80	&	3.38	&	25	&	3.00	&	3.48	&	20	&	5.12	&	4.39	&	6	\\
7136	&	[\tinyion{Ar}{III}]	&	$-$0.391	&	10.95	&	8.48	&	15	&	8.24	&	7.64	&	25	&	5.79	&	4.95	&	6	\\
7320+30	&	[\tinyion{O}{II}]	&	$-$0.413	&	8.49	&	6.58	&	20	&	$\cdots$	&	$\cdots$	&	$\cdots$ 	&	$\cdots$	&	$\cdots$	&	$\cdots$	\\
7320	&	[\tinyion{O}{II}]	&	$-$0.410	&	$\cdots$	&	$\cdots$	&	$\cdots$	&	$\cdots$	&	$\cdots$	&	$\cdots$	&	1.27	&	1.08	&	12	\\
7330	&	[\tinyion{O}{II}]	&	$-$0.411	&	$\cdots$	&	$\cdots$	&	$\cdots$	&	$\cdots$	&	$\cdots$	&	$\cdots$	&	1.17	&	0.99	&	12	\\
7753	&	[\tinyion{Ar}{III}]	&	$-$0.452	&	$\cdots$	&	$\cdots$	&	$\cdots$	&	$\cdots$	&	$\cdots$	&	$\cdots$	&	1.37	&	1.14	&	12	\\
\hline
\end{tabular}
\end{table*}
}

\begin{equation}
\begin{split}
\frac{I(\lambda)}{I({\rm H}\beta)} = \frac{F(\lambda)}{F({\rm H}\beta)} 10^{f(\lambda)c({\rm H}\beta)} & \left(1-\frac{EW_{\it abs}(\lambda)}{EW(\lambda)}\right) \\
 \times & \left(1-\frac{EW_{\it abs}({\rm H}\beta)}{EW({\rm H}\beta)}\right)^{-1}\frac{100.0}{\eta},
\label{eq:01}
\end{split}
\end{equation}
where $F(\lambda)$ is the the absolute flux for each line, $EW(\lambda)$ is the equivalent width observed in $\lambda$, $EW_{\it abs}(\lambda)$ is the theoretical equivalent width with respect to $EW_{\it abs}({\rm H}\beta)$, and $\eta$ is a value very close to 100.00 and represents a small correction to $I({\rm H}\beta)$. There are two things to note here, the first one, related to the theoretical values of the $I$(\ion{H}{I}), is that there are small variations on the theoretical values with temperature and density, however we can not determine a proper $T_e$ and $n_e$ without knowing the line intensities; to be sure to use the correct values, from an initial value of $T_e$ and $n_e$ we determine the line intensities, and with these we determine $T_e$ and $n_e$; we iterate this procedure until it converges, this is standard procedure. The second, related to the overall normalization, is not a standard procedure, instead of normalizing all the lines to $I({\rm H}\beta)=100.0$; we decided to normalize the reddening-corrected fluxes to the Balmer decrement as a whole; this often results in having H$\beta$ intensities slightly different than 100.0 (but within 1$\sigma$), in this case $\eta$. The decision to present a value of $I({\rm H}\beta)$ different from 100.0, comes from the desire to have the best possible determination of the abundances; for example, when we measure the N$^{+}$ we use $I(6548)$ and $I(6584)$ simultaneously to include all the available [\ion{N}{II}] photons; for the same reason we do the same for the \ion{H}{I} photons. In this work $I({\rm H}\beta) = 100.0$ represents the best guess at the real value of $I({\rm H}\beta)$, based on 8 Balmer lines, and not only on the observed value of $I({\rm H}\beta)$.

Tables \ref{tab:lines1} and \ref{tab:lines2} show: in column (1) the adopted laboratory wavelength, $\lambda$, in column (2) the identification for each line, column (3) the extinction law value used for each line \citep{Seaton1979}, columns (4) $-$ (6), and (7) $-$ (9) represent the flux $F(\lambda)$, the intensity $I(\lambda)$, and the percentage error for the intensity (the error includes: calibration, photon count, reddening, and underlying absorption uncertainties), respectively for each object.

\section{Physical conditions}

\subsection{Temperature and Density \label{sec:tem_den}}

Temperatures and densities have been determined with the line diagnostics presented in Tables \ref{tab:lines1} and \ref{tab:lines2}, most of them (all except for $n$[\ion{Fe}{III}]) using the package \texttt{PyNeb} (version 1.1.13) developed by \citet{pyneb2015}.

For the temperature we used two zone models. We estimated the temperature of the low-excitation zone, $T_{\it low}$, using: $T$[\ion{N}{II}] $(\lambda6548+\lambda6383)/\lambda5755$, $T$[\ion{O}{II}] $(\lambda3726+\lambda3729)/(\lambda7320+\lambda7330)$, and $T$[\ion{S}{II}] $(\lambda6717+\lambda6731)/(\lambda4069+\lambda4076)$, and the temperature of the high-excitation zones, $T_{\it high}$, using: $T$[\ion{O}{III}] $(\lambda4959+\lambda5007)/\lambda4363$ only. The exception to this is UM~420~A, where no low excitation temperature was measured; for this object we take advantage of the correlation between $T$[\ion{O}{II}] and $T$[\ion{O}{III}] \citep[see, for example][]{Izotov1994,Perez-Montero2009}; these correlation suggests that at $12,500\,$K the temperatures are very similar, and thus we assume $T_{\it low}=T$[\ion{O}{III}]. Overall the temperatures of 7 regions lie in the $9,500<T_{\it low}<14,200\,$K and $11,600<T_{\it high}<15,400\,$K ranges; all moderately high temperatures, seeking to avoid the complications inherent to the hottest regions ( $15,000\lesssim T_{\it high} \lesssim 20,000\,$K, extremely low metallicity) as well as the complications produced for larger chemical evolution corrections (associated with intermediate to high metallicity and with lower temperatures: $log({\rm O/H})+12\lesssim 8.3$, $T_{\it high} \lesssim 12,000\,$K.

All the densities were in the low density limit, frequently with error bars larger than the determinations themselves, therefore we determined a single density to represent both zones. We were able to determine $n$[\ion{O}{II}] $\lambda3726/\lambda3729$ and the $n$[\ion{S}{II}] $\lambda6717/\lambda6731$. With such low densities it was meaningless to try to determine the [\ion{Cl}{III}] and [\ion{Ar}{IV}] densities; their error bars are at least an order of magnitude larger than the expected density. The other density at our disposal is the $n$[\ion{Fe}{III}] $I(\lambda4986)/I(\lambda4658)$ density, from the computations by \citet{Keenan2001}. This ratio is strongly dependent on density, going from $I(4986)/I(4658)\approx1.0$ at $n_{e}=100$ cm$^{-3}$ to $I(4986)/I(4658)\approx0.05$ at $n_{e}=3000$ cm$^{-3}$.

\begin{table*}
\caption{Temperatures and Densities for the sample. \label{tab:physic}}
\centering
\begin{tabular}{lccccccc}
\hline
  & \multicolumn{3}{c}{UM~160} & \multicolumn{3}{c}{UM~420} & TOL~0513$-$393 \\
  & A & B & C & A & B & C & \\
\hline
Temperature [K]  &  &  &  &  &  &  & \\
\null$T$[\ion{N}{II}]  &  $15\,000\pm3\,000$& $11\,000\pm1\,500$& $8\,000\pm1\,200$  & $\cdots$& $9\,800:$ & $\cdots$ & $15\,000\pm2\,500$ \\
\null$T$[\ion{O}{II}]  &  $12\,000\pm650$ & $12\,000\pm750$ & $11\,000\pm450$ & $\cdots$&$\cdots$ &  $\cdots$ & $16\,300\pm1\,000$ \\
\null$T$[\ion{S}{II}]  &  $14\,000\pm2\,200$& $9\,000\pm900$   & $8\,800\pm700$   & $\cdots$ & $11\,300\pm1\,200$ & $10\,200\pm1\,700$ & $11\,300\pm1\,300$ \\
$T_{\it low}$       & $13\,500\pm2\,000$ & $10\,700\pm900$ & $9\,500\pm700$ & $12\,500\pm1\,600\,^{\rm a}$ & $11\,300\pm1\,200$ & $10\,200\pm1\,700$ & $14\,200\pm1\,300$ \\
\null$T$[\ion{O}{III}]$\;= \,T_{\it high}$ &  $14\,000\pm450$ & $13\,200\pm420$ & $11\,900\pm400$ & $12\,500\pm1\,600$ & $11\,600\pm450$ & $14\,300\pm800$ & $15\,400\pm280$ \\
\hline
Density [cm$^{-3}$]  &  &  &  &  &  &  & \\
\null$n$[\ion{O}{II}]   &$<70$ & $<100$ & $<60$ & $<130$ & $100\pm30$ & $<50$  & $150\pm40$ \\
\null$n$[\ion{S}{II}]   & $100\pm60$ & $<200$ & $40:$  & $<250$ &$150\pm70$ & $100:$ & $300\pm130$ \\
\null$n$[\ion{Cl}{III}] & $\cdots$ & $\cdots$ & $<1\,200$ & $\cdots$ & $\cdots$ & $\cdots$ & $<3\,500$ \\
\null$n$[\ion{Fe}{III}] & $55\pm33$ & $173\pm70$ & $158\pm68$ & $\cdots$ & $56\pm33$ & $70\pm40$ & $100\pm30$ \\
$n_{\it adopted}$   & $60\pm40$ & $150\pm70$ & $100\pm75$ & $<200$ & $90\pm50$ & $75\pm50$ & $120\pm40$ \\
\hline
\multicolumn{8}{l}{$^{\rm a}$ Derived from $T$[\ion{O}{III}], see text.}
\end{tabular}
\end{table*}

\subsection{Temperature Fluctuations \label{sec:t2}}

We obtain the ionic abundance using two different methods: a) the direct method, and b) the $t^2$ formalism \citep{Peimbert1967}. The $t^2$ formalism supposes the temperature in the \ion{H}{II} region is inhomogeneous in the observed volume \citep[e.g.,][]{Peimbert1967,Peimbert1969,Peimbert2012}.

Temperature inhomogeneities (or the ADF) affect strongly the abundance determinations for heavy elements, this in turn will affect the $\Delta Y/\Delta Z_O$ correction to determine $Y_{\rm P}$. This should not be a very big correction for two reasons: first at higher temperatures the effect of thermal inhomogeneities tends to be smaller; second for very low metallicities, even a correction of a factor of two,  will be small. The bigger problem for the $Y_{\rm P}$ determination is a difference on the temperature: RLs are brighter on cooler regions, while CELs are brighter on warmer regions; the diagnostics we obtained on section \S~\ref{sec:tem_den} are all from CELs and therefore reflect the temperatures present on the warmer regions, while the H and He RLs represent slightly cooler regions (regions approximately 1500K cooler than what was used in section \S~\ref{sec:tem_den}); this temperature difference can result on  over-estimations of $Y$ and $Y_{\rm P}$ of up to 2\% \citep[e.g.][]{Peimbert2000,Peimbert2007}.

Unfortunately there is no direct determination of a RLs temperature, or of any ADF for this set of objects. To select the $t^2$ value, required for both the ADF correction and the $T_{\it CEL}-T_{\it RL}$ correction, we used the value recommended by \citet{Peimbert2012}. The $t^2$ value depends on the specific characteristics of each object, according to the degree of ionization of the objects in our sample, we use the value of $t^2=0.029\pm0.004$, which corresponds to zone IIa of figure 4 of \citet{Peimbert2012}.

\section{Chemical abundances}

\subsection{Heavy element ionic abundances}

We used \texttt{PyNeb} \citep{pyneb2015} to  estimate the chemical abundances through the direct method. We adopted two-zone models characterized by the degree of ionization: low-ionization, where $T_{\it low}$ is the mean of $T_e$([\ion{N}{II}]) and $T_e$([\ion{O}{II}]) for singly ionized heavy elements, and high-ionization, where $T_{\it high}$ is $T_e$([\ion{O}{III}]) for multiple ionized elements.

In addition, we estimated the ionic abundances considering thermal inhomogeneities with $t^2=0.029\pm0.004$; these were obtained using the correction to the direct method abundances estimated by the $t^2$ formalism \citep{Peimbert1967,Peimbert1969}:
\begin{eqnarray}
\left[\frac{n_{\it CEL}(X^{+i})}{n({\rm H}^{+})}\right]_{t^{2}\neq0.00}& = & \frac{T({\rm H}\beta)^{\alpha}T(\lambda_{nm})^{0.5}}{T_{(4363/5007)}^{\alpha+0.5}} \nonumber \\
 && \times \, \exp\left[-\frac{\Delta E_{n}}{kT_{(4363/5007)}}+\frac{\Delta E_{n}}{kT(\lambda_{nm})}\right] \nonumber \\
 && \times \left[\frac{n_{\it CEL}(X^{+i})}{n({\rm H}^{+})}\right]_{t^{2}=0.00},
\end{eqnarray}
where $\alpha=-0.89$ is the temperature dependence of H$\beta$, $\Delta E_{n}$ is the difference of energy between the excited level of the CEL and the ground level, and $T(\lambda_{nm})$ and $T({\rm H}\beta)$ are line temperatures, as described by \citet{Peimbert1967}.

\subsection{Helium ionic abundances}

To compute the He$^{+}$ abundance, we use the {\tt Helio14} code, which is described in \citet{Peimbert2012}. The code determines the most likely values for He$^+$/H$^+$, $n_e$(\ion{He}{I}), $\tau(2^3S)$, and $t^2$(\ion{He}{I}); it uses as input a set of \ion{He}{I}/H$\beta$ line intensity ratios along with their uncertainties. Finally, it compares the theoretical ratios to the observed ones minimizing $\chi^2$. The nature of the code requires line intensities that represent the gaseous phase; similarly to what we did for the underlying stellar absorption for \ion{H}{I} lines, it is important to consider the correction of the underlying stellar absorption for \ion{He}{I} lines; this absorption is overall about 5 to 20 times smaller for the \ion{He}{I} lines, but since these lines are about 10 times fainter than the \ion{H}{I} lines it is approximately as important; unlike for the \ion{H}{I} lines, the correction is quite different for each \ion{He}{I} line; we use the results by \citet{Gonzalez1999} to correct lines with $\lambda<5000$\AA\ and the results by \citet{Peimbert2005} to correct lines redder than $5876$\AA. The He$^{+}$/H$^{+}$ abundances are presented in Table \ref{tab:ion-ab}.

To determine the He$^{++}$ abundance, we use the recombination \ion{He}{II} $\lambda4686$ line and the recombination coefficients given by \citet{Storey1995}. Table \ref{tab:ion-ab} shows the He$^{++}$/H$^{+}$ abundance in each \ion{H}{II} region.

\begin{table*}
\centering
\caption{Ionic abundances for each \ion{H}{II} region. \label{tab:ion-ab}}
\begin{tabular}{lcccccc}
\hline
Ion  & $t^{2}=0.000$  & $t^{2}=0.029\pm0.004$ & $t^{2}=0.000$ & $t^{2}=0.029\pm0.004$ & $t^{2}=0.000$  & $t^{2}=0.029\pm0.004$ \\
\hline
& \multicolumn{2}{c}{UM~160~A} & \multicolumn{2}{c}{UM~160~B} &\multicolumn{2}{c}{UM~160~C} \\
\hline
He$^{+}$&$10.898\pm0.015$&$10.896\pm0.015$&$10.891\pm0.015$&$10.889\pm0.015$&$10.903\pm0.014$&$10.901\pm0.014$\\
He$^{++}$&$8.599\pm0.121$&$8.599\pm0.121$&$9.298\pm0.057$&$9.298\pm0.057$&$\cdots$&$\cdots$\\
N$^{+}$&5.74$\pm$0.06&5.79$\pm$0.06&6.03$\pm$0.08&6.31$\pm$0.09&6.24$\pm$0.05&6.49$\pm$0.06\\
O$^{+}$&7.30$\pm$0.09&7.36$\pm$0.09&7.54$\pm$0.11&7.86$\pm$0.12&7.77$\pm$0.08&8.06$\pm$0.09\\
O$^{++}$&7.82$\pm$0.04&7.86$\pm$0.04&7.86$\pm$0.04&8.06$\pm$0.05&7.89$\pm$0.04&8.08$\pm$0.05\\
Ne$^{++}$&7.05$\pm$0.05&7.09$\pm$0.05&7.16$\pm$0.05&7.37$\pm$0.06&7.21$\pm$0.05&7.40$\pm$0.06\\
S$^{+}$&5.44$\pm$0.06&5.49$\pm$0.06&5.62$\pm$0.07&5.91$\pm$0.08&5.82$\pm$0.05&6.07$\pm$0.06\\
S$^{++}$&6.03$\pm$0.08&6.07$\pm$0.08&6.04$\pm$0.09&6.25$\pm$0.09&6.23$\pm$0.05&6.42$\pm$0.06\\
Ar$^{++}$&5.53$\pm$0.04&5.57$\pm$0.04&5.71$\pm$0.04&5.90$\pm$0.05&5.67$\pm$0.04&5.84$\pm$0.05\\
Ar$^{+3}$&4.58$\pm$0.09&4.61$\pm$0.09&$\cdots$&$\cdots$&$\cdots$&$\cdots$\\
Cl$^{++}$&4.27$\pm$0.10&4.31$\pm$0.10&4.32$\pm$0.11&4.53$\pm$0.11&4.43$\pm$0.09&4.61$\pm$0.09\\
\hline
&\multicolumn{2}{c}{UM~420~A} & \multicolumn{2}{c}{UM~420~B} & \multicolumn{2}{c}{UM~420~C} \\
\hline
He$^{+}$ &$11.067\pm0.051$&$11.065\pm0.051$&$10.942\pm0.015$&$10.940\pm0.015$&$10.944\pm0.020$&$10.942\pm0.020$\\
He$^{++}$&$\cdots$&$\cdots$&$8.723\pm0.092$&$8.723\pm0.092$&$\cdots$&$\cdots$\\
N$^{+}$&6.62$\pm$0.27&6.80$\pm$0.27&6.50$\pm$0.43&6.52$\pm$0.43&6.13$\pm$0.11&6.15$\pm$0.11\\
O$^{+}$&7.69$\pm$0.40&7.91$\pm$0.40&7.72$\pm$0.16&7.77$\pm$0.16&7.39$\pm$0.11&7.39$\pm$0.11\\
O$^{++}$&7.73$\pm$0.38&8.07$\pm$0.39&7.93$\pm$0.08&7.97$\pm$0.08&7.77$\pm$0.15&7.82$\pm$0.15\\
Ne$^{++}$&7.23$\pm$0.45&7.61$\pm$0.46&7.27$\pm$0.10&7.28$\pm$0.10&7.09$\pm$0.18&7.13$\pm$0.18\\
S$^{+}$&6.01$\pm$0.24&6.23$\pm$0.25&5.85$\pm$0.11&6.33$\pm$0.13&5.64$\pm$0.10&5.65$\pm$0.10\\
S$^{++}$&$\cdots$&$\cdots$&6.38$\pm$0.17&6.39$\pm$0.17&6.05$\pm$0.32&6.09$\pm$0.32\\
Ar$^{++}$&$\cdots$&$\cdots$&5.69$\pm$0.17&5.74$\pm$0.17&5.47$\pm$0.32&5.53$\pm$0.32\\
Cl$^{++}$&$\cdots$&$\cdots$&4.27$\pm$0.33&4.31$\pm$0.33&$\cdots$&$\cdots$\\
\hline
&\multicolumn{2}{c}{TOL~0513$-$393} & & & & \\
\cline{1-3}
He$^{+}$&$10.924\pm0.010$&$10.922\pm0.010$&\\
He$^{++}$&$8.918\pm0.057$&$8.918\pm0.057$&\\
N$^{+}$&5.50$\pm$0.08&5.60$\pm$0.08&\\
O$^{+}$&6.64$\pm$0.12&6.75$\pm$0.12&\\
O$^{++}$&7.89$\pm$0.02&7.98$\pm$0.02&\\
Ne$^{++}$&7.11$\pm$0.02&7.20$\pm$0.02&\\
S$^{+}$&5.03$\pm$0.07&5.12$\pm$0.07&\\
S$^{++}$&5.90$\pm$0.02&5.99$\pm$0.02&\\
Ar$^{++}$&5.27$\pm$0.03&5.35$\pm$0.03&\\
Ar$^{+3}$&5.10$\pm$0.03&5.19$\pm$0.03&\\
Cl$^{++}$&4.02$\pm$0.09&4.11$\pm$0.09&\\
\cline{1-3}
\multicolumn{3}{c}{In units of $12+\log n(X^{+i})/n({\rm H^{+}})$.}
\end{tabular}
\end{table*}

\subsection{Heavy element total abundances}

In general, we can not observe all ionization stages for the elements present in our objects. To obtain the total abundances, we adopt commonly used ionization correction factors, ICFs, from the literature \citep[i.e.][]{Peimbert1969,Stasinska1978, PerezM2007, Delgado2014}.

To obtain the oxygen abundance we used
\begin{equation}
{\rm ICF}({\rm O}^{+}+{\rm O}^{++}) = \left[1-\frac{n({\rm He}^{++})}{n({\rm He})}\right]^{-1}.
\end{equation}
\citep{Peimbert2000}

To obtain the nitrogen and neon abundances, we used ICFs by \citet{Peimbert1969}:
\begin{equation}
{\rm ICF}({\rm N}^{+}) = \frac{n({\rm O}^{+})+n({\rm O}^{++})+n({\rm O}^{+3})}{n({\rm O}^{+})}
\end{equation}
and
\begin{equation}
{\rm ICF}({\rm Ne}^{++}) = \frac{n({\rm O}^{+})+n({\rm O}^{++})+n({\rm O}^{+3})}{n({\rm O}^{++})}.
\end{equation}

The sulphur abundance was obtained using ICF proposed by \citet{Stasinska1978}:
\begin{equation}
{\rm ICF}({\rm S}^{+}+{\rm S}^{++}) = \left[1-\left(1-
\frac{n({\rm O}^{+})}{n({\rm O})}\right)^{3}\right]^{1/3}.
\end{equation}

For argon, we have used the ICF by \citet{PerezM2007}:
\begin{equation}
{\rm ICF}({\rm Ar}^{++}+{\rm Ar}^{3+}) = 0.928+0.364\left(1-x\right)+\frac{0.006}{1-x},
\end{equation}
where $x={n({\rm O}}^{++})/\left[n({\rm O}^{+})+n({\rm O}^{++})\right]$.

Finally, to estimate the chlorine abundance, we used a reimplementation of the ICF proposed by \citet{Delgado2014}:
\begin{equation}
{\rm ICF}({\rm Cl}^{++}) = 2.914 \left(1-\left[ \frac{n({\rm O}^{++})}{n({\rm O}^{+})+{n({\rm O}}^{++})} \right]^{0.21}\right)^{0.75}.
\end{equation}

We have obtained the total abundances using ionic abundances derived from both methods: the direct method and the $t^{2}$ formalism. Table \ref{tab:tot} shows both sets of results (for $t^{2}=0.00$ and for $t^{2}=0.029\pm0.004$).

\begin{table*}
\centering
\caption{Total abundances for each HII region. \label{tab:tot}}
\begin{tabular}{lcccccc}
\hline
Ion  & $t^{2}=0.000$  & $t^{2}=0.029\pm0.004$ & $t^{2}=0.000$ & $t^{2}=0.029\pm0.004$ & $t^{2}=0.000$  & $t^{2}=0.029\pm0.004$ \\
\hline
& \multicolumn{2}{c}{UM~160~A} & \multicolumn{2}{c}{UM~160~B} &\multicolumn{2}{c}{UM~160~C} \\
\hline
He& $10.900\pm0.015$ & $10.898\pm0.015$ & $10.902\pm0.015$ & $10.900\pm0.015$& $10.886\pm0.013$ & $10.884\pm0.013$ \\
O&7.93 $\pm$ 0.03&8.05 $\pm$ 0.04&8.03 $\pm$ 0.04&8.16 $\pm$ 0.05 &8.14 $\pm$ 0.04&8.31 $\pm$ 0.05 \\
N&6.38 $\pm$ 0.10&6.47 $\pm$ 0.10&6.52 $\pm$ 0.13&6.63 $\pm$ 0.13 &6.61 $\pm$ 0.09&6.75 $\pm$ 0.09 \\
Ne&7.17 $\pm$ 0.06&7.28 $\pm$ 0.07&7.33 $\pm$ 0.07&7.47 $\pm$ 0.07 &7.45 $\pm$ 0.08&7.62 $\pm$ 0.08 \\
S&6.01 $\pm$ 0.19&6.12 $\pm$ 0.19&6.06 $\pm$ 0.19&6.19 $\pm$ 0.19 &6.25 $\pm$ 0.37&6.40 $\pm$ 0.37 \\
Ar&5.69 $\pm$ 0.04&5.80 $\pm$ 0.04&5.88 $\pm$ 0.04&6.01 $\pm$ 0.05 &5.92 $\pm$ 0.05&6.07 $\pm$ 0.05 \\
Cl&4.25 $\pm$ 0.21&4.36 $\pm$ 0.21&4.34 $\pm$ 0.22&4.47 $\pm$ 0.22 &4.45 $\pm$ 0.50&4.60 $\pm$ 0.50 \\
\hline
&\multicolumn{2}{c}{UM~420~A} & \multicolumn{2}{c}{UM~420~B} & \multicolumn{2}{c}{UM~420~C} \\
\hline
He & $11.067\pm0.051$ &$11.065\pm0.051$ & $10.945\pm0.015$ & $10.943\pm0.015$ & $10.944\pm0.020$ & $10.942\pm0.020$ \\
O &8.01 $\pm$ 0.02&8.30 $\pm$ 0.03 &8.14 $\pm$ 0.07&8.15 $\pm$ 0.06&7.92 $\pm$ 0.06&7.96 $\pm$ 0.07 \\
N &6.94 $\pm$ 0.04&7.19 $\pm$ 0.04 &6.92 $\pm$ 0.10&6.97 $\pm$ 0.11&6.66 $\pm$ 0.13&6.72 $\pm$ 0.11 \\
Ne &7.51 $\pm$ 0.05&7.84 $\pm$ 0.07 &7.48 $\pm$ 0.04&7.47 $\pm$ 0.05&7.24 $\pm$ 0.07&7.27 $\pm$ 0.09 \\
S &6.05 $\pm$ 0.09&6.42 $\pm$ 0.11 &6.55 $\pm$ 0.05&6.96 $\pm$ 0.07&6.27 $\pm$ 0.05&6.33 $\pm$ 0.06\\
Ar &$\cdots$ &$\cdots$ &5.90 $\pm$ 0.07&5.93 $\pm$ 0.09&4.26 $\pm$ 0.13&5.67 $\pm$ 0.13 \\
Cl &$\cdots$ &$\cdots$ &4.44 $\pm$ 0.14&4.88 $\pm$ 0.12&$\cdots$&$\cdots$ \\
\hline
&\multicolumn{2}{c}{TOL~0513$-$393} & & & & \\
\cline{1-3}
He& $10.967\pm0.012$ & $10.965\pm0.012$   & \\
O&7.91 $\pm$ 0.02&8.00 $\pm$ 0.02  &\\
N&6.77 $\pm$ 0.13&6.85 $\pm$ 0.13  &\\
Ne&7.13 $\pm$ 0.04&7.22 $\pm$ 0.04  &\\
S&5.83 $\pm$ 0.36&5.92 $\pm$ 0.36  &\\
Ar&5.52 $\pm$ 0.02&5.60 $\pm$ 0.03  &\\
Cl&3.95 $\pm$ 0.46&4.04 $\pm$ 0.46  &\\
\cline{1-3}
\multicolumn{3}{c}{In units of $12+\log n(X)/n({\rm H})$.}
\end{tabular}
\end{table*}

\subsection{Total Helium abundance}

The main problem to compute the He abundance, is the possible presence of neutral helium within the \ion{H}{II} region; therefore, in principle, it is necessary to use an ICF(He$^+$+He$^{++}$). In the general case, there is not a good way to estimate the ICF(He$^+$+He$^{++}$). For objects with a low or medium ionization degree, the presence of neutral helium is important and the ${\rm ICF}({\rm He}^++{\rm He}^{++})>1.00$; for objects with high ionization degree the presence of neutral helium within the nebula can be small and the ICF can be close to unity (and in the extreme case of very high ionization degree, the ${\rm {ICF}({\rm He}^++{\rm He}^{++})}\lesssim1.00$).

Our sample includes only objects with high ionization degree, and we expect all ${\rm ICF}({\rm He}^++{\rm He}^{++})$ to be within $0.5\%$ of unity and we will assume ${\rm ICF}({\rm He}^++{\rm He}^{++})=1.00$. Table \ref{tab:tot} shows the results for each total abundance in each \ion{H}{II} region.

\section{Primordial helium abundance}

To determine $Y_{\rm P}$, we must first determine the fraction of helium by mass. We consider the normalization by unit mass given by $X + Y + Z = 1$; where $X$ represents hydrogen, $Y$ represents helium, and $Z$ represents the rest of the elements by mass. While we do know the He/H ratio, we do not know the $Y$ (nor the $Z$) value. However, $Z/X$ is directly correlated with the fraction of oxygen by mass, which is in turn correlated with the O/H ratio; we will assume that the oxygen by mass amounts to $55\%\pm10\%$ of the $Z$ value \citep{Peimbert2007}.

We use the helium and oxygen abundances (by number) presented in Table \ref{tab:tot} to compute $Y$ and $Z$ (by mass). To derive $Y_P$ from $Y$, we have to estimate the fraction of helium present in these objects due to galactic chemical evolution. We use the following equation:
\begin{equation}
Y_{\rm P}=Y-Z_O\,\frac{\Delta Y}{\Delta Z_O},
\end{equation}
where $Z_O$ is the fraction of $Z$ contributed by oxygen (the oxygen abundance by mass), and $\Delta Y/\Delta Z_O$ represents the ratio of the contributions of helium and oxygen to the ISM from stellar nucleosynthesis. We adopted $\Delta Y/\Delta Z_O  = 3.3\pm0.7$ from \citet{Peimbert2016}, this is very close to the $3.4\pm1.1$ value measured by \citet{Kurichin2021}.

In Table \ref{tab:errors}, we show the values obtained for each individual \ion{H}{II} region, as well as their errors; we have separated the errors, presenting first the sum of all the statistical errors, and then each individual systematic errors listed in \citet{Peimbert2007}. We isolate each systematic error in an effort to show the current limit of this technique and to show that an addition of a large sample of objects cannot decrease these errors, and in fact poorly chosen objects, could potentially increase the overall error bars. Also, in the last rows of Table \ref{tab:errors}, we present the average of the sample, first separating each individual systematic error, then with all the systematic errors added together, and finally with a single total error bar.

Overall the errors in our measurements will become part of our statistical errors (line intensities, temperatures, etc.), while the errors of our astronomical ``tools" will be part of our systematic errors (our reddening function, our chemical evolution correction, etc.). The reason why systematic errors are not uniform is because the systematic errors produced by reddening correction, chemical evolution, and collisional H excitation for any object (or set of objects) are not uniform; i.e. the effect of a given imperfection on our tool can give us errors of variable magnitude on our determinations (e.g. an error in the reddening function could make us over estimate He/H in all objects, but would have more impact in objects where the reddening correction is larger).

\begin{table*}
\begin{center}
\caption{Primordual helium abundances from our sample.}
\label{tab:errors}
\begin{tabular}{lccccccc}
\hline
                   &             & \multicolumn{5}{c}{errors} \\
\cline{3-8}
\ion{H}{II} region & $Y_{\rm P}$ & statistical$^{\rm a}$ & $f(\lambda)^{\rm b}$ & He r.c.$^{\rm c}$ & H r.c.$^{\rm d}$ & H col.ex.$^{\rm e}$ & $\Delta Y/Z_O$$^{\rm f}$ \\
\hline
UM160A  & $0.2370$ & $\pm0.0068$ & $\pm0.0010$ & $\pm0.0010$ & $\pm0.0005$ & $\pm0.0009$ & $\pm0.0008$\\
UM160B  & $0.2365$ & $\pm0.0069$ & $\pm0.0005$ & $\pm0.0010$ & $\pm0.0005$ & $\pm0.0005$ & $\pm0.0011$\\
UM160C  & $0.2351$ & $\pm0.0058$ & $\pm0.0010$ & $\pm0.0010$ & $\pm0.0005$ & $\pm0.0004$ & $\pm0.0016$\\
UM420A  & $0.3122$ & $\pm0.0288$ & $\pm0.0005$ & $\pm0.0010$ & $\pm0.0005$ & $\pm0.0007$ & $\pm0.0015$\\
UM420B  & $0.2554$ & $\pm0.0072$ & $\pm0.0007$ & $\pm0.0010$ & $\pm0.0005$ & $\pm0.0005$ & $\pm0.0010$\\
UM420C  & $0.2584$ & $\pm0.0099$ & $\pm0.0004$ & $\pm0.0010$ & $\pm0.0005$ & $\pm0.0008$ & $\pm0.0008$\\
TOL 0513-393 & $0.2496$ & $\pm0.0049$ & $\pm0.0004$ & $\pm0.0010$ & $\pm0.0005$ & $\pm0.0010$ & $\pm0.0008$\\
\hline
Mean  & $0.2448$ & $\pm0.0027$ & $\pm0.0007$ & $\pm0.0010$ & $\pm0.0005$ & $\pm0.0007$ & $\pm0.0010$\\
      & \multicolumn{2}{l}{$0.2448\pm0.0027\pm0.0018$} \\
      & \multicolumn{2}{l}{$0.2448\pm0.0033$} \\
\hline
\end{tabular}\\
\end{center}
\begin{flushleft}
\hspace{2.5cm}$^{\rm a}$ The sum of all the statistical errors.\\
\hspace{2.5cm}$^{\rm b}$ Systematic error due to imperfect reddening law, large $c({\rm H}\beta)$ are more affected.\\
\hspace{2.5cm}$^{\rm c}$ Systematic error associated to the theoretical intensities of the He recombination lines.\\
\hspace{2.5cm}$^{\rm d}$ Systematic error associated to the theoretical intensities of the H recombination lines.\\
\hspace{2.5cm}$^{\rm e}$ Systematic error due to the imperfect understanding of the collisional excitation of the Balmer lines. \\
\hspace{2.5cm}$^{\rm f}$ Systematic error associated to the chemical evolution $\Delta Y / \Delta Z_O$
\end{flushleft}
\end{table*}

In Table \ref{tab:yp}, we compare our $Y_{\rm P}$ determination with recent \ion{H}{II} region $Y_{\rm P}$ determinations in the literature. As comparison, the result by the \citet{Planck2020}, corresponds to $Y_{\rm P}=0.24714\pm0.00050$. It is clear that the $Y_{\rm P}$ values derived by different groups provide an important constraint to the Big Bang theory. Most determinations are dominated by systematic errors rather than statistical ones (of course, it is relatively easy to decrease the statistical errors increasing the number of objects), on the other hand there is a lower limit to the systematic error of the best objects of about $\pm0.0018$ (this value can be slightly larger for very hot objects, for objects with high metallicity, or for heavily reddened objects). Our error bar of $\pm0.0033$ is relatively close to this systematic limit.

\begin{table}
\centering
\caption{Primordial helium abundance, $Y_{\rm P}$ reported in the literature. \label{tab:yp}}
\begin{tabular}{lc}
\hline
$Y_{\rm P}$ source & $Y_{\rm P}$           \\
\hline
\citet{Izotov2014}    &  $0.2551\pm0.0022$ \\
\citet{Aver2015}      &  $0.2449\pm0.0040$ \\
\citet{Peimbert2016}  &  $0.2446\pm0.0029$ \\
\citet{Valerdi2019}   &  $0.2451\pm0.0026$ \\
\citet{Vital2019}     &  $0.243 \pm0.005$  \\
\citet{Aver2020}      &  $0.2453\pm0.0034$ \\
\citet{Hsy2020}       &  $0.2436\pm0.0040$ \\
\citet{Kurichin2021}  &  $0.2462\pm0.0022$ \\
This work             &  $0.2448\pm0.0033$ \\
\hline
\citet{Planck2020}     & $0.24714^{+(0.00012)0.00049}_{-(0.00013)0.00049}$  \\
\hline
\end{tabular}
\end{table}

Following the discussion of section 8 by \citet{Valerdi2019} the number of neutrino families from this new determination is $2.91\pm0.22$ alternatively the expected neutron mean lifetime amounts to $872\pm17\,$s. Our result of the number of neutrino families, is consistent with previous works \citep[e.g][]{Aver2015,Aver2020,Peimbert2016,Vital2019,Valerdi2019,Hsy2020,Kurichin2021}, and agree with the presence of three neutrino families, which is consistent with laboratory determinations \citep[e.g.][]{Aleph2006,Pattie2018}.

Moreover, our result of the neutron mean lifetime, together with the results obtained from previous works \citep[e.g][]{Aver2015,Aver2020,Peimbert2016,Vital2019,Valerdi2019,Hsy2020,Kurichin2021}, are within $1\sigma$ from the value $\tau_n=879.4\pm0.6\,$s presented by the \citet{2020PTEP.2020h3C01P} from laboratory determinations.

\begin{figure}
\includegraphics[width=\columnwidth]{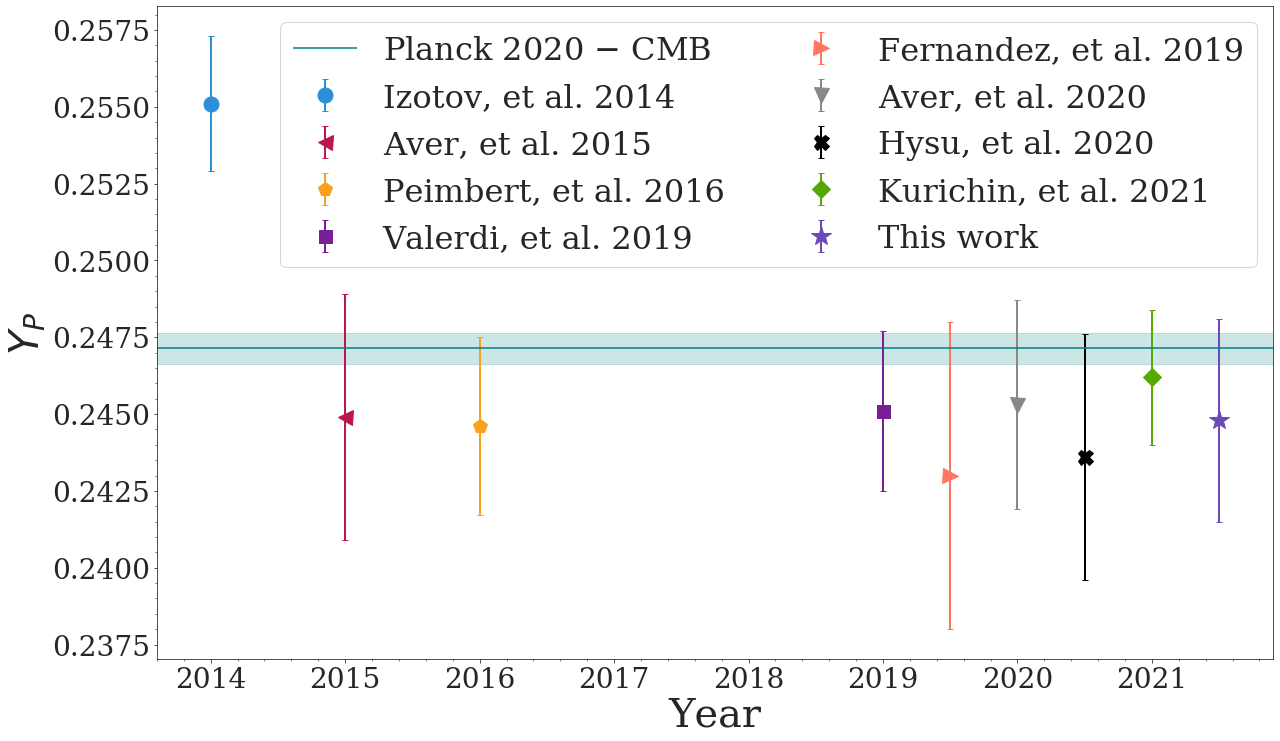}
\caption{Recent $Y_{\rm P}$ measurements based on \ion{H}{II} regions.}
\label{fig:history}
\end{figure}

In Figure \ref{fig:history}, we show the determinations of the primordial helium abundance using \ion{H}{II} regions from 2014 to the present. The $Y_{\rm P}$ values derived by different groups provide an important constraint to the Big Bang theory. The main source of error is due to systematic uncertainties. In addition, we include the result derived by \citet{Planck2020} from: the cosmic microwave background (CMB) data, dark energy + cold dark matter ($\Lambda$CDM) cosmological models, and standard big bang nucleosynthesis (SBBN) models.

We can see that, other than the work by Izotov \citep[that could be an overestimation of the primordial helium abundance;][]{Kurichin2021}, all other $Y_{\rm P}$ determinations derived from \ion{H}{II} regions fall a little bit under the Planck Collaboration determination. This likely means that the differences are not statistical. Although these $\sim0.0020$ differences are consistent with nominal systematic errors, some of the systematic errors raised since \citet{Peimbert2007} are associated with individual samples and not the field in general (i.e. each author used different characteristics in the determination of $Y_{\rm P}$ such as: different extinction law, collisional contribution to the H lines, $\Delta Y/\Delta Z_O$ value, etc). Thus, while this difference can be explained by a systematic effect in the field of \ion{H}{II} regions, the possibility that it comes from the determinations of the SBBN (implying that the Big Bang is not standard and that there are small discrepancies in our understanding of physical constants or there is an ingredient missing in the BBN calculations) cannot be ruled out.

\section{Summary and Conclusions}

We studied a sample of 7 extragalactic \ion{H}{II} regions in three metal-poor galaxies. We follow the standard analysis to determine physical conditions and chemical abundances; we also include an analysis using the $t^2$ formalism. We focused our work on the helium abundance, to obtain such abundance we use a code dedicated to helium in a reliable way ({\tt Helio14} code).

Once the helium abundance is obtained, we correct by $\Delta Y/\Delta Z_O$ and estimate the primordial helium abundance, $Y_{\rm P}$. We first obtained $Y_{\rm P}$ for each of the objects, separating the statistical and systematic errors, and then we averaged all these determinations to obtain a single determination of $Y_{\rm P}$, minimizing the final error. Our final primordial helium abundance is $Y_{\rm P}=0.2448\pm0.0033$, which is consistent with most of the previous estimates obtained using low metallicity \ion{H}{II} regions, as well as with the determination using the CMB, in \citet{Planck2020}.

Our $Y_{\rm P}$ determination allows us to estimate the number of neutrino families, $N_{\nu}=2.91\pm0.22$ and the neutron half-life value $\tau_n=872\pm17$s. Both results are consistent with laboratory determinations \citep[e.g.,][]{Aleph2006,Pattie2018}.

When comparing $Y_{\rm P}$ values determined from \ion{H}{II} regions with those derived from the CMB we find a systematic difference between both sets of determinations; while the difference is within the error bars for each determination, the fact that the last 7 published determinations of $Y_{\rm P}$ from \ion{H}{II} regions fall bellow the CMB determination suggests these differences arise from the systematic errors associated with  \ion{H}{II} region $Y_{\rm P}$ determinations, but without ruling out the possibility that it comes from the BBN $Y_{\rm P}$ determinations.

As a final note: we want to improve the accuracy of the $Y_{\rm P}$ determinations. This will require more and better spectra, it will require to look for objects with smaller than average contribution to the systematic errors; but, at this point, it also requires better determinations of the atomic emission of He$^0$ and even a newer study on the extinction law. If we can achieve all these, it will increase the reliability of determining the physical parameters of \ion{H}{II} regions as well as of $Y_{\rm P}$.

\section*{Acknowledgements}

We acknowledge UNAM-DGAPA for supporting this project through the grant PAPIIT IG100319. We would also like to acknowledge an anonymous referee for a carefully reading and many helpful suggestions.


\vspace{0.5cm}
\begin{center}
{\bf Data Availability.} {\it The data underlying this article will be shared on reasonable request to the corresponding author.}
\end{center}


\bibliographystyle{mnras}
\bibliography{references.bib}



\bsp	
\label{lastpage}
\end{document}